\documentclass[article]{aa} 

\usepackage{txfonts}
\usepackage[utf8]{inputenc}
\usepackage{graphicx}
\usepackage{natbib}
\bibpunct{(}{)}{;}{a}{}{,} 
\usepackage{caption}
\usepackage{color}
\usepackage{subcaption}
\usepackage{float}
\usepackage{wasysym}
\usepackage{array}
\usepackage{hyperref}
\usepackage{textcomp}
\usepackage{comment}

\usepackage{txfonts}
\begin{document}
\title{Influence of the C/O ratio on titanium and vanadium oxides in protoplanetary disks}
\authorrunning{Ali-Dib et al.}

\author{M.~Ali-Dib\inst{1}, O.~Mousis\inst{1}, G.~S.~Pekmezci\inst{2}, J.~I.~Lunine\inst{3}, N.~Madhusudhan\inst{4} and J.-M.~Petit\inst{1}}
\institute{Institut UTINAM, CNRS-UMR 6213, Observatoire de Besan\c con, BP 1615, 25010 Besan\c{c}on Cedex, France
\email{mdib@obs-besancon.fr}
\and 
Dipartimento di Astronomia, Universita' di Roma Tor Vergata, Via della Ricerca Scientifica 1, 00133 Roma, Italy
\and
Center for Radiophysics and Space Research, Space Sciences Building, Cornell University, Ithaca, NY 14853, USA
\and
Department of Physics and Department of Astronomy, Yale University, New Haven, CT 06511, USA
}

\date{Received 26/04/2013; accepted 12/11/2013}

  \abstract
{The observation of carbon-rich disks have motivated several studies questioning the influence of the C/O ratio on their gas phase composition in order to establish the connection between the metallicity of hot-Jupiters and that of their parent stars.}
{We to propose a method that allows the characterization of the adopted C/O ratio in protoplanetary disks independently from the determination of the host star composition. Titanium and vanadium chemistries are investigated because they are strong optical absorbers and also because their oxides are known to be sensitive to the C/O ratio in some exoplanet atmospheres.}
{We use a commercial package based on the Gibbs energy minimization technique to compute the titanium and vanadium equilibrium chemistries in protoplanetary disks for C/O ratios ranging from 0.05 to 10.  Our calculations are performed for pressures in the 10$^{-6}$--10$^{-2}$ bar domain, and for temperatures ranging from 50 to 2000 K.}
{We find that the vanadium nitride/vanadium oxide and titanium hydride/titanium oxide gas phase ratios strongly depend on the C/O ratio in the hot parts of disks ($T\ge$ 1000 K). Our calculations suggest that, in these regions, these ratios can be used as tracers of the C/O value in protoplanetary disks.
}
{}


\maketitle

\section{Introduction}
The recent detection of carbon-rich planets (hereafter CRPs), with C/O ratios $\geq$ 1 in their envelopes \citep{2011Natur.469...64M}, have stimulated research on their physical properties and the scenarios that may lead to their formation \citep{2011ApJ...743..191M,2011ApJ...743L..16O,2012ApJ...751L...7M}. Carbon-rich disks have also been observed in the last years \citep{2006Natur.441..724R} and motivated several studies on the influence of the C/O ratio on their gas phase composition \citep{2008ApJ...681.1624M,2012ApJ...757..192J}. These studies suggested that, as the C/O ratio increases in the gas phase of the disk, all the available O goes into organics, CO, CO$_2$ and CH$_3$OH, so that the gas phase becomes H$_2$O-free and the remaining C is in the form of CH$_4$, enabling formation conditions of a CRP. In the case of hot Jupiters the determination of this ratio is possible due to the extremely high temperatures of their upper atmospheres.  This ratio critically influences the relative concentrations of several spectroscopically dominant species. For example, H$_2$O and CH$_4$ abundances\footnote{Fractional abundance with respect to H$_2$} can vary by several orders of magnitude if C/O is modified by factors from 2 to 4 \citep{2012ApJ...758...36M}. 

The formation scenario of CRPs is still poorly understood, in particular when considering the recent observation of WASP 12b, a CRP orbiting a carbon-poor star, leading to the conclusion that the disk C/O ratio might be different from that of the host star \citep{2011Natur.469...64M}. The properties of the protoplanetary disk needed to form the CRP WASP 12b have been investigated by \cite{2011ApJ...743..191M}. These authors retrieved the composition of the protoplanetary disk from that of planetesimals accreted during WASP 12b's formation and needed to match the observed volatile abundances in the planet's atmosphere. They concluded that the C/O ratio of 1 observed in WASP 12b requires a substantial oxygen depletion in the disk (factor of $\sim$0.4). The same approach was used by \cite{2012ApJ...751L...7M} to propose the formation of Jupiter through the accretion of condensed volatiles in the cold outer part of an oxygen-depleted primordial nebula. This scenario reproduces the measured Jovian elemental abundances at least as well as the hitherto canonical model of Jupiter formed in a disk of solar composition. The resulting O abundance in Jupiter's envelope is then moderately enriched by a factor of $\sim$2 $\times$ solar (instead of $\sim$7 $\times$ solar) and is found to be consistent with values predicted by thermochemical models of the atmosphere. This model suggests that water ice might have been distributed inhomogeneously beyond the snowline in the primordial nebula. Alternatively, it has been proposed that the envelopes of CRPs could be formed from an oxygen-depleted gas from the nebula. In this case, the oxygen depletion would result from the water condensation and incorporation at earlier epochs in the building blocks of the planetary cores \citep{2011ApJ...743L..16O}. However, this scenario predicts that the abundances of carbon, nitrogen, and other ultravolatiles are solar in the envelopes of Hot-Jupiters and is not found consistent with the supersolar abundances measured at Jupiter. Interestingly, another scenario proposed by \cite{2004ApJ...611..587L} suggests that Jupiter or CRPs could have formed in a zone of the disk where carbonaceous matter dominates, rather than water ice. All these models and observations outline a non trivial relation between the C/O ratios of the host stars, their protoplanetary disks and eventually their planets. For this reason it is important to establish a reliable probe of this ratio. 

Here we suggest that the sampling of the abundances of titanium-- and vanadium--bearing minerals could allow to probe the C/O ratio in protoplanetary disks. Titanium oxide (TiO, TiO$_2$, etc) and vanadium oxide (VO, VO$_2$, etc) are strong optical absorbers. They were first observed at optical wavelengths in M-type stars \citep{1962ApJ...136...21M}. Since then, these molecules have been detected in various astrophysical environments. TiO and TiO$_2$ have been observed at sub-millimeter wavelengths in the circumstellar envelope of VY Canis Majoris \citep{2013A&A...551A.113K}. Moreover, TiO and VO have been identified around protostars in optical emission bands \citep{2012AJ....143...37H} and also in the infrared emission spectrum of S-type star atmospheres \citep{2012A&A...543L...2S}. Recently, TiO and VO have been proposed to be the cause of the thermal inversions observed in several highly irradiated hot-Jupiters \citep{2003ApJ...594.1011H,2008ApJ...678.1419F} in which the determination of the C/O ratio is possible \citep{2012ApJ...758...36M}. However, it has been found that an implausible eddy coefficient value is needed for the occurrence of the TiO/VO induced thermal inversion \citep{2009ApJ...699.1487S,2010ApJ...720.1569K}. The effect of C/O on the abundances of TiO and VO in the atmospheres of hot Jupiters has been investigated by \cite{2012ApJ...758...36M} who found that C/O = 1 leads to severe depletions in TiO and VO in the atmospheres of CRPs. On the other hand, a recent HST observation of the hot Jupiter WASP 19b shows that this planet has no or low levels of TiO with a moderate C/O ratio, suggesting that this lack of observable TiO is possibly due to rainout or breakdown from stellar activity \citep{2013MNRAS.434.3252H}. Another recent study also reports that all stars with TiO emission exhibit a low C/O ratio \citep{2012A&A...543L...2S}. Given the similarities that exist between the atmospheres of giant planets and the gas phase of protoplanetary disks (some common temperature and pressure ranges, H$_2$-dominated compositions, dynamical effects), the influence of the C/O ratio on Ti and V chemistries in disks deserves to be investigated.  

In Section 2, we present the chemical model used to relate TiO and VO abundances to C/O for a large range of temperatures and pressures. Section 3 is devoted to the presentation of our major results and to their comparison with observations. Results obtained for some important volatile species are also included. Conclusions are given in Section 4.


 
\section{Computing the chemistries of gas and solid phases in protoplanetary disks}

In order to obtain the disk solid and gaseous composition, we use the HSC chemistry commercial package. It is based on the Gibbs energy minimization method, originally developed by \cite{1958JChPh..28..751W}. In this method, the calculation of phase equilibrium is made by minimizing the Gibbs free energy of the system, at constant temperature $T$ and pressure $P$, with respect to the number of moles of each component in each phase $n^k_i$. For a system with $NP$ phases and $NC$ components, we have

\begin{equation} 
G=\sum\limits_{k=1}^{NP}\sum\limits_{i=1}^{NC} n^k_i \mu^k_i,
\end{equation} 

\noindent where $n^k_i$ and $\mu^k_i$ are the number of moles and the chemical potential of component $i$ in phase $k$, respectively. The chemical potential is a function of the composition of phase $k$ at temperature $T$ and pressure $P$ \citep{rossi}. {This method for equilibrium chemistry calculations assumes local thermodynamical equilibrium.} Several codes, based on this approach, have been developed in the last decades. The most known are SOLGASMIX \citep{1990ApJS...72..417S}, NASA's CEA\footnote{http://www.grc.nasa.gov/WWW/CEAWeb/ceaHistory.htm} and the HSC Chemistry package developed by Outotec Research\footnote{http://www.outotec.com/en/Products--services/HSC-Chemistry}. In this work we opted to use the HSC Chemistry package because its built-in database of minerals is the largest existing one. It contains for example $\sim$75 Ti bearing minerals.

In our equilibrium calculations, usually one favors a long list of elements to get the most precise results possible, and to avoid any possible hidden non linear effect from an ignored molecule. For this reason, we chose to use the entire HSC Chemistry database of elements relevant to solar elemental abundances totalizing 1224 gases and solids (excluding C$_n$H$_m$ for n$>$3). Once the initial list of elements and their respective abundances have been defined (see Table \ref{abun}), we have selected the set of molecular species that form from these elements (with null initial abundances). We made the reasonable assumption that the gas phase composition of the disk under study has the same elemental composition as that of the host star. Here, all our calculations are based on the solar photosphere abundances taken from \cite{2009ARA&A..47..481A}.
\footnote{$\dagger$ For C/O = 1, we used O/H$_2$ = C/H$_2$ = 5.90 $\times 10^{-4}$.}

\begin{table}[h]
\centering \caption{Initial solar elemental abundances}
\begin{tabular}{ccc}
\hline 
\hline 
\multicolumn{1}{c}{Element}
 			&	\multicolumn{1}{c}{Solar abundance in Kmol}
 			&	\multicolumn{1}{c}{X/H$_2$} 							\\
\hline
H 			& $1.00\times10^9$		&								\\
He 			& $9.55\times10^7$		&$1.90\times10^{-1}$				\\
N 			& $7.41\times10^4$		&$1.48\times10^{-4}$				\\
O			& $5.37\times10^5$  	&$1.07\times10^{-3}\dagger$				\\
C			& $2.95\times10^5$		&$5.90\times10^{-4}$				\\
Na			& $1.91\times10^3$		&$3.80\times10^{-7}$				\\
Mg			& $4.37\times10^4$		&$4.60\times10^{-8}$				\\
Al 			& $3.09\times10^3$		&$6.00\times10^{-6}$				\\
Ni			& $1.88\times10^3$		&$3.60\times10^{-7}$				\\
Si			& $3.55\times10^4$		&$7.00\times10^{-5}$				\\
P			& $2.82\times10^2$		&$5.60\times10^{-7}$				\\
S			& $1.45\times10^4$		&$2.90\times10^{-6}$				\\
Ca			& $2.40\times10^3$		&$4.80\times10^{-6}$				\\
Ti			& $9.77\times10^1$		&$1.94\times10^{-7}$				\\
Cr			& $4.79\times10^2$		&$9.40\times10^{-7}$				\\
Fe			& $3.47\times10^4$		&$6.80\times10^{-5}$				\\
V			& $9.33\times10^0$		&$1.86\times10^{-8}$				\\
\hline
\end{tabular}
\label{abun}
\end{table}

\begin{figure*}[ht]
\begin{center}
\resizebox{\hsize}{!}{\includegraphics[angle=0]{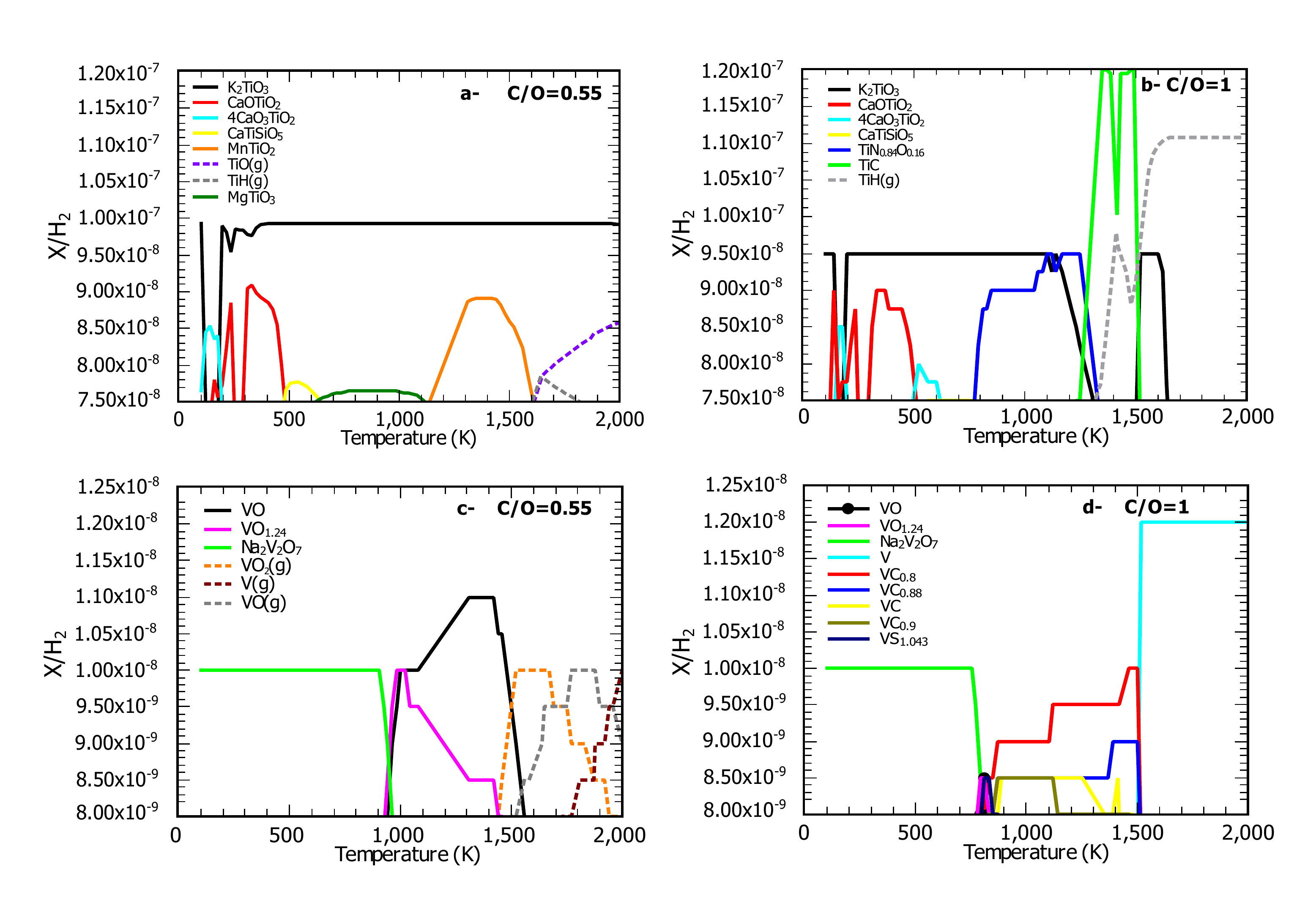}}
\caption{(a) Abundances of titanium-bearing solid and gaseous compounds computed at equilibrium at a disk's pressure of 10$^{-4}$ bar as a function of the temperature for C/O = 0.55. Solid lines are for solid species and dashed lines for gases. (b) Same as (a) but for C/O = 1. (c) Abundances of vanadium-bearing solid and gaseous compounds computed at equilibrium at a disk's pressure of 10$^{-4}$ bar as a function of the temperature for C/O = 0.55. (d) Same as (c) but for C/O = 1.}
\label{Ti1}
\end{center}
\end{figure*}

\begin{figure}[h]
\begin{center}
\resizebox{\hsize}{!}{\includegraphics[angle=0]{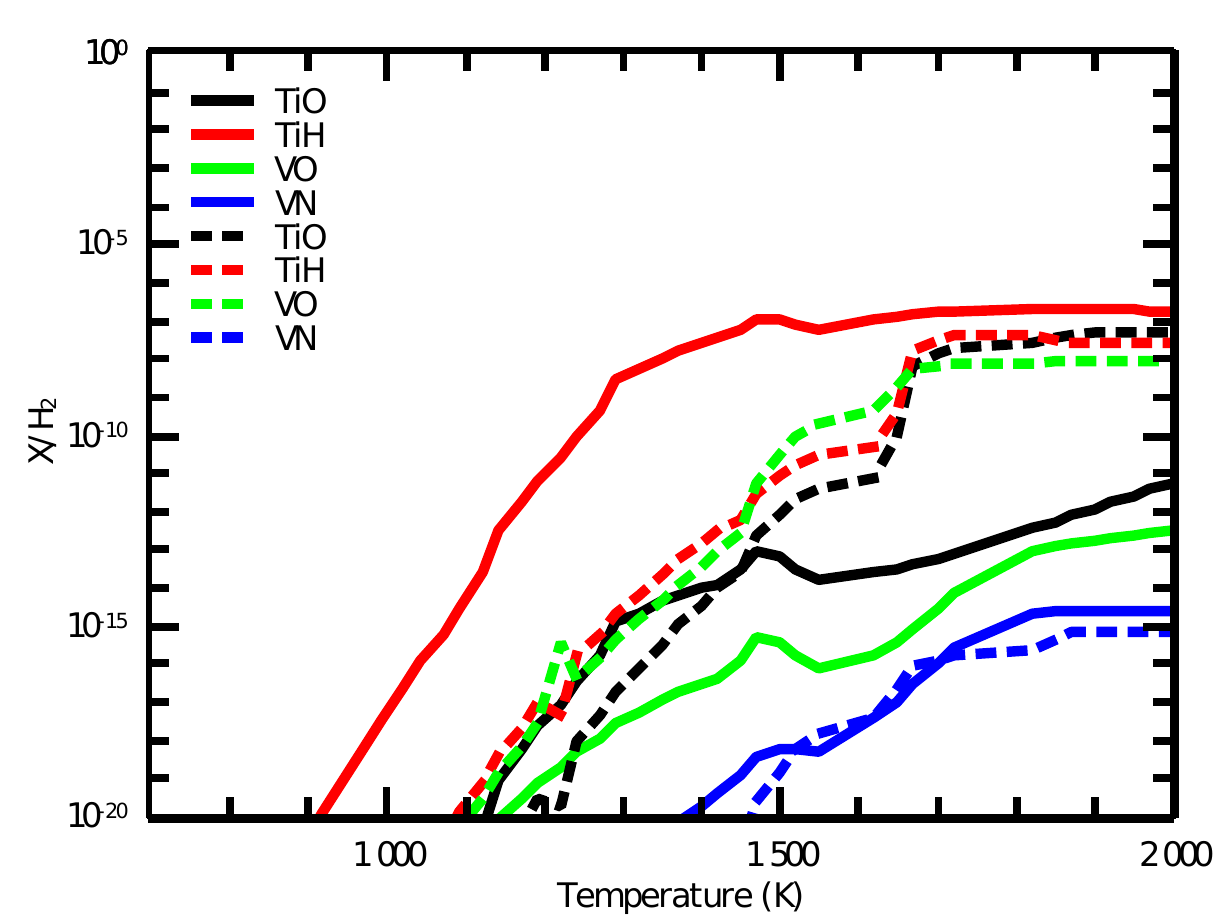}}
\caption{Gas phase equilibrium abundances of TiO, TiH, VO and VN computed at a disk pressure of 10$^{-4}$ bar for C/O = 0.55 (dashed lines) and C/O = 1 (solid lines).}
\label{Tin}
\end{center}
\end{figure}

\section{Results}
\subsection{Influence of C/O on Ti and V chemistries}

Figure \ref{Ti1} displays the abundances of solid and gaseous Ti and V bearing species, for C/O = 0.55 and 1. In order to get C/O = 1 in the disk's gas phase, we opted to decrease the initial abundance of oxygen \citep{2011ApJ...743..191M,2012ApJ...751L...7M}. For the sake of clarity, we just represented the abundances of species that are at most 10 times lower than the most abundant one, as more than 1,000 abundances of species are simultaneously computed. All our calculations are made for a total gas pressure of 10$^{-4}$ bar, which corresponds to the average pressure in the inner regions {( $\sim$ 1-2 AU )} of protoplanetary disks\citep{2005A&A...442..703H}\footnote{The details of the used disk model are presented in section 3.3 .}. For C/O = 0.55 and in the case of Ti chemistry, we note the dominance of the solid (s) mineral K$_2$TiO$_3$(s) at all temperatures except in the range 100-200 K where 4CaO$_3$TiO(s) is the most abundant. At temperatures higher than 1600 K, we note the existence of substantial amounts of the gaseous (g) compounds TiO(g) and TiH(g). In the case of V chemistry, Na$_2$V$_2$O$_7$(s) and VO(s) are the most abundant minerals at temperatures lower than $\sim$1500 K and the gaseous compounds VO$_2$(g), VO(g), and V(g) become dominant at higher temperatures. For C/O = 1, and in the case of Ti-bearing solids, K$_2$TiO$_3$(s) now dominates in the 1500--1600 K interval and also at temperatures lower than $\sim$1200 K, except in the 100--200 K interval where 4CaO$_3$TiO$_2$(s) is the most abundant species. TiC(s) and TiH(g) become the most abundant solid species in the 1200--1500 K and 1650--2000 K temperature ranges, respectively. In the case of V-bearing solids, Na$_2$V$_2$O$_7$(s) becomes dominant only for temperatures lower than $\sim$800 K. Solid VO(s) now exists at temperatures centered around $\sim$800 K and its maximum abundance is 4 times less than in the first case. At this value of C/O, we note that the abundances of gaseous  VO$_2$(g), VO(g) and TiO(g) are $\sim$ one thousand times lower than in the former case, making them not visible in the panels of Fig. \ref{Ti1}. Our calculations then suggest that Ti and V oxides present in the gas phase are orders of magnitudes more abundant in the solar C/O case than in the C/O~=~1 case. It should be noted that Ti and V bearing minerals were studied in a large number of meteorites \citep{2008ApJ...682.1450N,1997M&PS...32..231R,2006ApJ...647L..37L,2007GeCoA..71.3098S}. 

Figure \ref{Tin} shows the evolution of the gas phase abundances of TiO, TiH, VO and VN computed as a function of temperature for C/O = 0.55 and 1 in the disk. The abundances of these species increase with the growth of temperature at both C/O values. One can note that the abundances of TiO and VO are significantly higher at lower C/O values and at high temperatures. On the othe hand, TiH is more abundant at high C/O irrespective of temperature and the abundance of VN only weakly depends on the C/O value. 

\begin{figure}[h]
\begin{center}
\resizebox{\hsize}{!}{\includegraphics[angle=0]{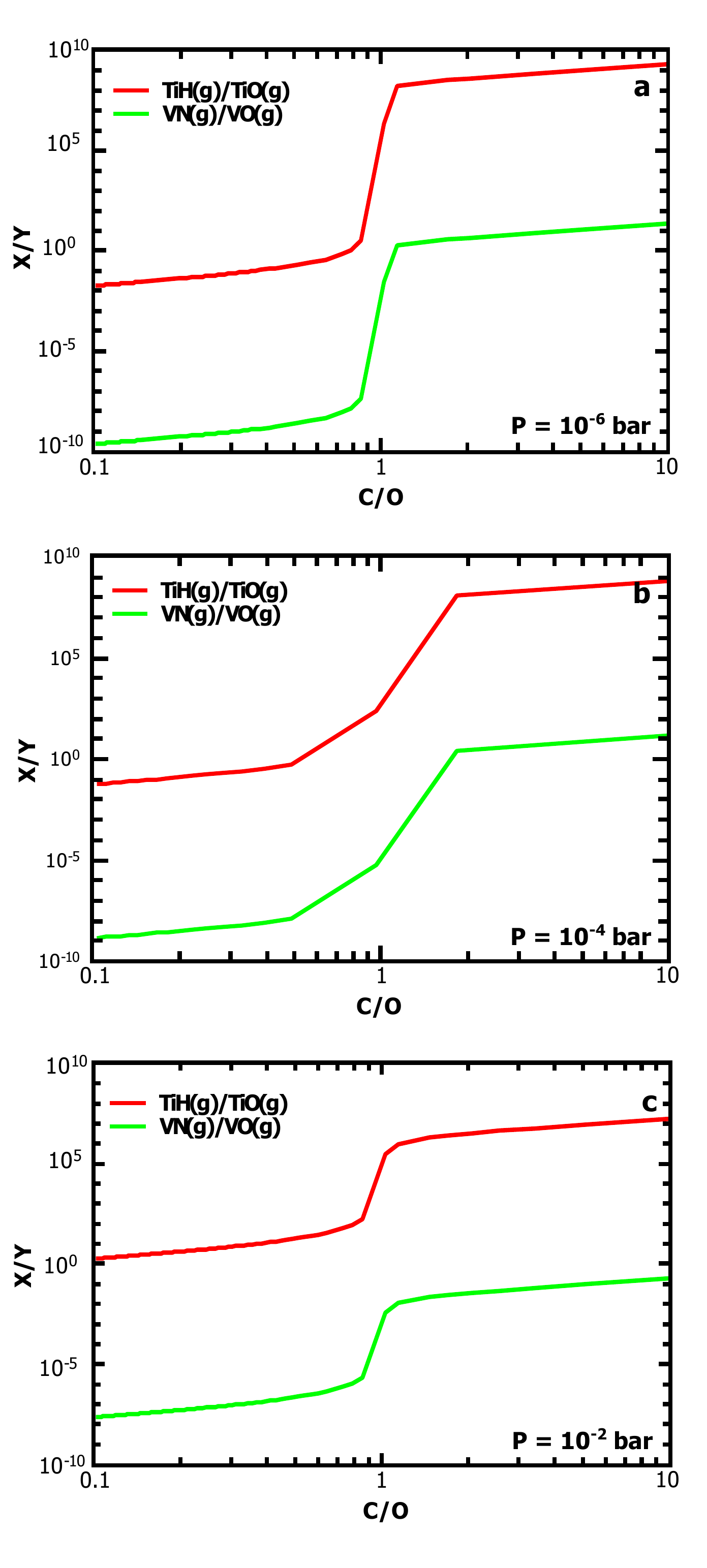}}
\caption{VN/VO$_2$ and TiH/TiO gas phase ratios as a function of C/O in protoplanetary disks for pressures ranging from 10$^{-6}$ to 10$^{-2}$ bar {and for $T$ = 1700 K.}} 
\label{Ti2}
\end{center}
\end{figure}

Figure \ref{Ti2}b represents the VN/VO and TiH/TiO gas phase ratios computed as a function of C/O ranging between 0.1 and 10 in protoplanetary disks. As mentioned above, the C/O ratio is set to the desired value by varying the oxygen abundance. Computations have been conducted at 1700 K and at disk pressures of $10^{-6}$, $10^{-4}$ and $10^{-2}$ bar. These molecules have been selected because their abundance ratios heavily depend on the value of the adopted C/O ratio. Gaseous VN also exists in the same temperature range as VO but its abundance is too low (by $\sim$4-5 orders of magnitudes in both C/O cases) to make it visible in Fig. \ref{Ti1}. At the considered pressures, there is a strong dependence of VN/VO and TiH/TiO gas phase ratios with the adopted value of C/O ratio. For example, an increase of the C/O ratio from 0.3 to 2 induces a steep increase of the TiH/TiO and VN/VO ratios by $\sim$10 orders of magnitudes at P = $10^{-4}$ bar, respectively. Between C/O = 0.1 and 0.3 and beyond C/O = 2, the two molecular ratios increase slightly linearly.

Our calculations suggest that these ratios can be used as tracers of the C/O ratio in protoplanetary disks. It is worth mentioning that calculations similar to those presented above have been done by varying the C abundance and fixing the O abundance, i.e. the opposite approach, in order to investigate the existence of any hidden non linear effect. The results obtained were found almost identical to those presented above, with a slight increase in the abundances of C bearing species for C/O$>$0.5. One should note that among the four species, VN has the lowest abundance for solar C/O, making its detection more critical than the other molecules.

\subsection{Influence of the disk's pressure}

\begin{figure*}[ht]
\begin{center}
\resizebox{\hsize}{!}{\includegraphics[angle=0]{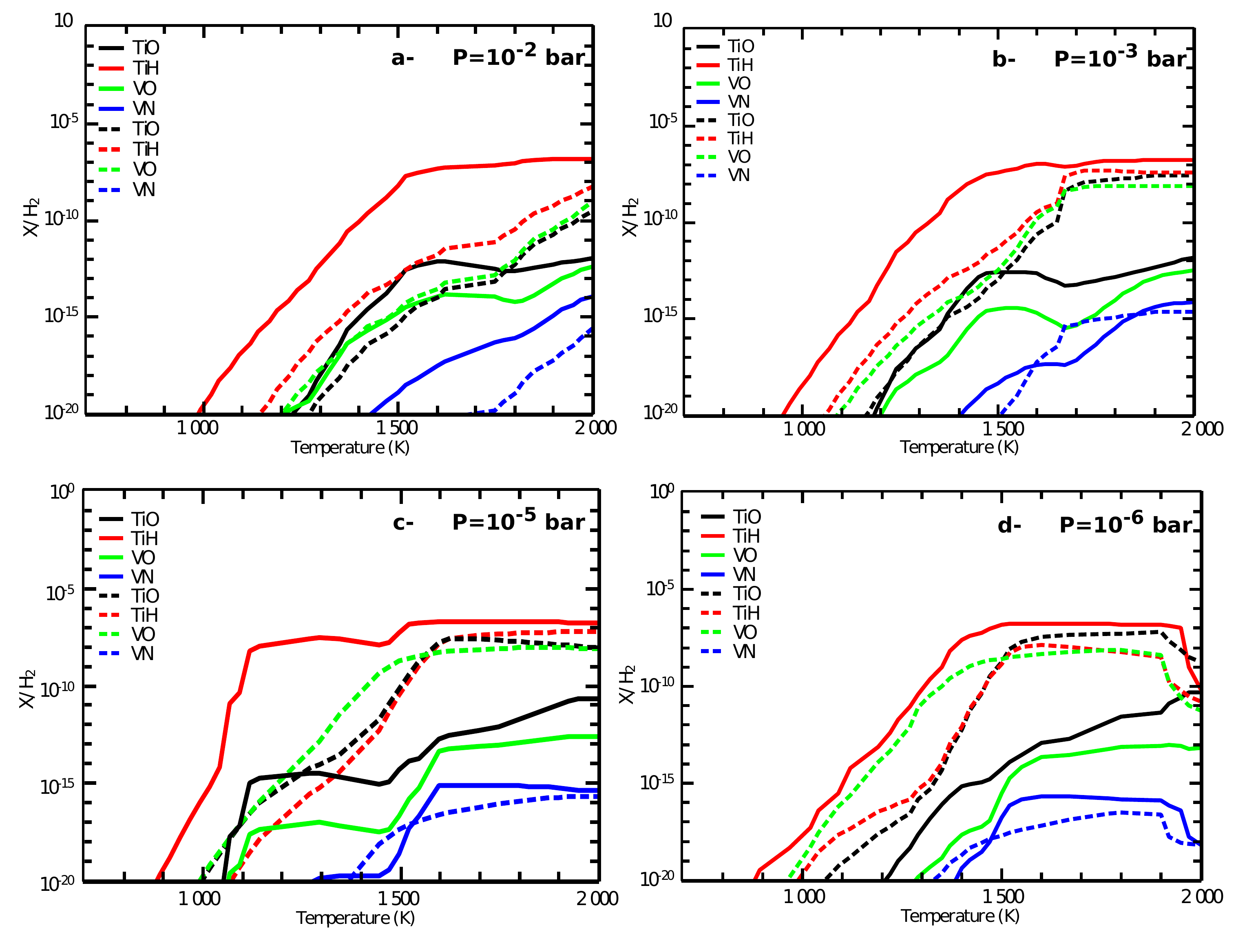}}
\caption{Abundances of gaseous TiO, TiH, VO and VN for total disk pressures of $10^{-2}$ bar (a), $10^{-3}$ bar (b), $10^{-4}$ bar (c), $10^{-6}$ bar (d). Solid and dashed lines represent the C/O = 1 and 0.55 cases, respectively.}
\label{pressure}
\end{center}
\end{figure*}

To quantify the effect of pressure on the Ti and V chemistries, we performed the same calculations as above but for total disk pressures in the $10^{-6}$--$10^{-2}$ bar range, and for C/O = 0.55 and 1. Results for TiO, TiH, VO and VN are presented in Figure \ref{pressure}. The main conclusion is that a variation of pressure over several orders of magnitude induces some non linear effects on the abundances of the considered species. Despite the fact that it is difficult to describe clear trends for these variations, we note that at $P$~=~$10^{-2}$ bar and for C/O = 0.55, the abundances of all four gaseous species are depleted by several orders of magnitude with respect to $10^{-3}$ bar, except for the very high temperatures ($\sim$1900--2000 K). This depletion is balanced by an important increase in the abundance of solid TiO$_2$. At the same pressure, we also note that the abundance of TiO is more important for C/O~=~1 than for C/O = 0.55 at disk temperatures $\le$~1800 K. This last effect is also present at $P$~$\sim$~$10^{-4}$--$10^{-3}$ bar and temperatures $\le$~1500 K, but is absent at $10^{-6}$ bar. At this latter pressure, we also note a decrease in the abundances of most of species over two orders of magnitude at C/O = 0.55, balanced by an increase in the abundances of solid minerals (mostly FeTiO$_3$). This decrease is restricted to the abundances of TiH and VN at C/O = 1.
Figure \ref{pressure} shows that the pressure variation can affect the abundances of TiO, TiH, VO and VN but it is not certain that a significant pressure drop can alter their detection. As mentioned below, TiO is detectable in some circumstellar environments with pressures of $\sim$10$^{-4}$ bar and the abundance of this species computed at 10$^{-6}$ bar is quite close.

We also computed the TiH/TiO and VN/VO ratios as a function of pressure. The results are presented in Figure \ref{Ti2}a-c. The same general trend as in Figure \ref{Ti2}b is found for the other investigated pressures, but with some slight differences. In particular, with increasing disk pressure, the TiH/TiO and VN/VO ratios increase by 2 orders of magnitude for C/O ratios $\ll$ 1. On the other hand, the ratios will decrease by 2 orders of magnitude with increasing disk pressure for C/O ratios $\gg$ 1.

\subsection{Comparison with observations}
In this section, we are going to investigate the observability of TiO, and compare our model to observations.

Using Spitzer infrared data, \cite{2012A&A...543L...2S} derived a TiO column density of $\sim$ $10^{17.25}$cm$^{-2}$ in the circumstellar environment of the S star NP Aurigae at temperatures higher than 1900 K. \cite{2013A&A...551A.113K} also reported the observation of TiO and TiO$_2$ at sub-millimetric wavelengths in the stellar environment of VY Canis Majoris, with column densities of $\sim$ $10^{14}$cm$^{-2}$ for both species, but for $T$ = 1000 $\pm$ 870 K. On the other hand, our calculations predict TiO abundances of $\sim$1.0 $\times10^{-8}$ and $\sim$1.0 $\times10^{-11}$ for respectively $T$ $\sim$1800 K and $T$ $\sim$1600 K, at $P$ $\sim$ $10^{-4}$ bar (typical gas pressure in circumstellar environments) in the C/O = 0.5 case (see Figure \ref{Tin}). For a typical scale height $H$ $\sim$5.0 $\times$ $10^{-3}$ AU, we find column densities $N_{1800 {\rm K}}^{TiO}=\frac{X_{\rm TiO}\times H}{KT}$ $\sim$ $2.9\times10^{17}$ cm$^{-2}$ and $N_{1600 {\rm K}}^{TiO}$ $\sim$ $2.9\times10^{14}$ cm$^{-2}$ for the computed TiO abundance. $N_{1800{\rm K}}$ is almost equal to the column density observed by \cite{2012A&A...543L...2S} and $N_{1600{\rm K}}$ is very close to the value derived by \cite{2013A&A...551A.113K}. Hence, it should be possible in principle to observe TiO in the hot inner disk.{ The TiO$_2$ abundance in our calculations is $\sim$ twice lower than that of TiO\footnote{The reason why it is not visible in Fig. \ref{Ti1}.} for T $\sim$ 1600 K, a result also compatible with \cite{2013A&A...551A.113K}.} 

\textcolor{black}{Following the referee's request, we also compared our equilibrium calculations to observations of volatile species in protoplanetary disks. To calculate the column densities, we used the one-dimensional $\alpha$-disk model of \cite{2005A&A...442..703H} to derive the thermodynamic parameters (temperature and scale height) needed. The model follows the evolution of a cloud-disk-star system with the following initial conditions (see \cite{2005A&A...442..703H} for details): $\alpha$ = 0.01, $M_{cloud}$ = 1 $M_\odot$, $\Omega_{cloud}$ = 3.0$\times$ $10^{-14}$ s$^{-1}$, $T_{cloud}$ = 10 K, $M_{0,star}$ = 0.1 $M_\odot$ and $T_{star}$ = 4000 K. We chose the model at 10$^{5}$ years.} 

\begin{figure*}[ht]
\begin{center}
\resizebox{\hsize}{!}{\includegraphics[angle=0]{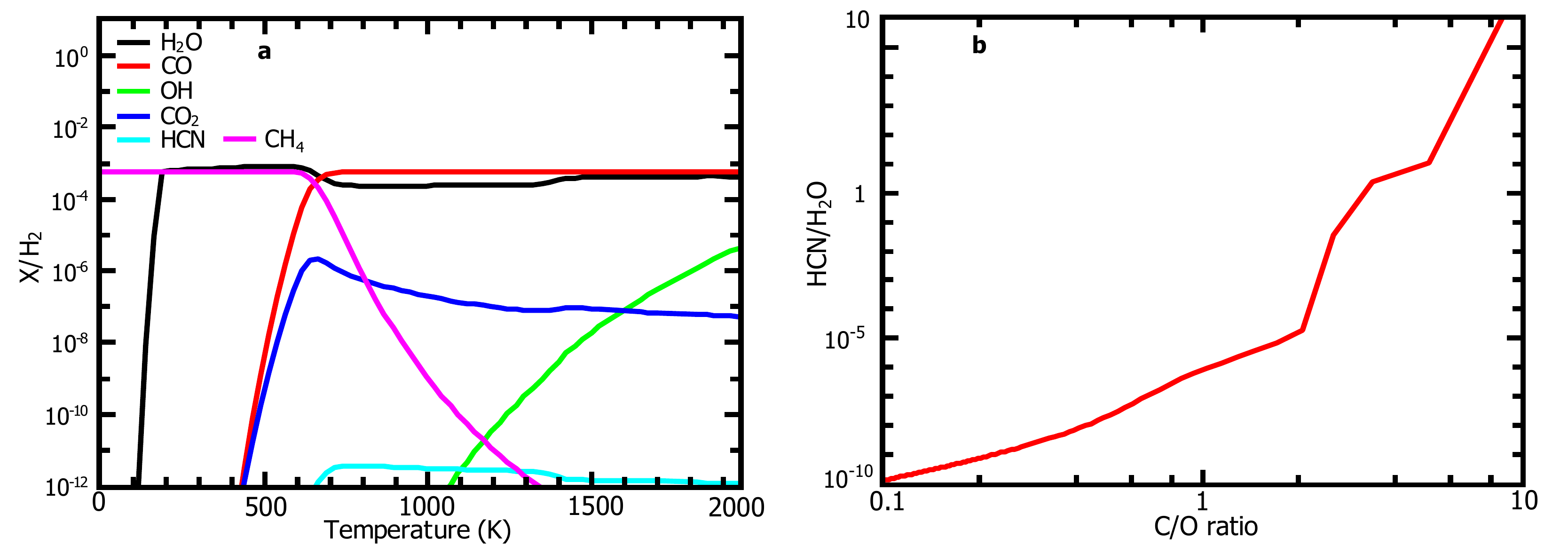}}
\caption{(a) Fractional abundances of selected volatile species {for C/O = 0.55} as a function of temperature (b) HCN/H$_2$O ratio as a function of the C/O ratio. All calculations were done at P = 10$^{-4}$ bar {for $T$ = 700 K.} }
\label{volatiles}
\end{center}
\end{figure*}

\cite{2011ApJ...733..102C} determined an average column density for HCN of $\sim$6~$\times$~$10^{16}$~cm$^{-2}$ at $T$ $\sim$700 K from Spitzer observations of inner regions of six different T-Tauri disks. However, this value is model dependent since it is defined by \cite{2011ApJ...733..102C} as the value averaged from two different (optically thin and optically thick) disk models. In the case of the optically thin disk model, these authors found $N$ $\sim$4~$\times$~$10^{15}$~cm$^{-2}$, which is one order of magnitude lower than the average column density. 
Since HCN is observed at $\sim$700 K, we used the a disk pressure of $\sim$ $10^{-3}$ bar and $H$ = 6$\times$10$^{-2}$ AU as calculated using the employed disk model. The calculated HCN abundance {for C/O = 0.55} is found to be $\sim 6\times$10$^{-11}$, corresponding to a column density of $\sim$4.2$\times$10$^{14}$ cm$^{-2}$, a value within an order of magnitude from observations for thin disks. Furthermore, we computed the abundance ratio C$_2$H$_2$/HCN, which is found to be $\sim5\times 10^{-5}$ at these disk conditions. This ratio is lower by several orders of magnitude compared to the value found by \cite{2011ApJ...733..102C} (around $10^{-2}$ to $10^{-1}$). \textcolor{black}{However, this important difference might be due to non-equilibrium effects (chemical kinetics, photochemistry) not taken into account in our model. For the sake of completeness, we also presented in Fig. \ref{volatiles}a the {fractional abundances with respect to H$_2$} of some major volatiles (H$_2$O, CO, CO$_2$, CH$_4$ and OH) that have been observed in disks and that are believed to be important in the planetary formation process. Another interesting observable for disk evolution and planetary formation is the HCN/H$_2$O ratio, since \cite{2011ApJ...733..102C} proposed that {it is correlated to the value of C/O ratio in T-Tauri disks.} In order to illustrate this correlation, we computed the HCN/H$_2$O ratio as a function of C/O in disks. The results, displayed in Fig. \ref{volatiles}b, show that the HCN/H$_2$O ratio increases as a function of the C/O ratio, with a particularly sharp slope for C/O$\geq$2. Finally, it is worth mentioning that our calculations performed in the C/O = 1 case are consistent with the conclusions of \cite{2006ApJ...647L..37L} who finds that TiC (observed in meteorites) is formed naturally at equilibrium under these conditions.}

\section{Conclusions and prospects}

We have performed computations of equilibrium chemistry describing the fate of Ti-- and V--bearing species in protoplanetary disks as a function of the C/O ratios, and using the Gibbs energy minimization method. This allowed us to find that the VN/VO and TiH/TiO gas phase ratios strongly depend on the degree of the C/O ratio in the hot parts of disks. Gaseous TiO and VO have been detected at optical wavelengths in the 1000--2000 K range in stellar envelopes and photospheres \citep{2012AJ....143...37H,2012A&A...543L...2S} at pressure regimes very close to those encountered in the inner part of protoplanetary disks \citep{2004A&A...417..751A}.

The results presented in this study are based on equilibrium calculations. We opted to neglect the influence of photochemistry and of turbulent diffusion for the following reasons: the proposed observations will be mainly in millimeter-wavelength, probing deep in the midplane of the disk, away from the surface photochemically active zone. The effects of turbulent diffusion are however less straightforward. Since we did not find any kinetic data concerning the gas phase chemistries of Ti- and V-bearing species, we have not been able to quantify the quenching effects caused by the different chemical and dynamical timescales if any. Experimental studies of the kinetic properties of the Ti- and V-bearing species will be needed to investigate further these effects.
 
Finally, observations of gaseous TiO and VO in disks must be spatially resolved in order to eliminate any confusion from stellar emissions. Such a high resolution should be attained with the new generation sub-millimetric instruments such as ALMA, from which several spatially resolved observations of different types of disks have been recently reported with angular resolutions sufficient to resolve the inner disks, where the molecules considered in the present work are observable \citep{2008Ap&SS.313...15V,2007ApJ...665..478K}. However, the midplane of the inner disk might be inaccessible to these instruments if dust extinction is significant at submm/mm wavelengths.  Observations of TiO emission in the 10 micron spectral region from circumstellar environments around AGB stars will be more sensitive (by an order of magnitude\footnote{http://www.stsci.edu/jwst/science/sensitivity}), and with higher spatial resolution than, those of Spitzer. Detection or a better upper limit by JWST  of TiO$_2$ and other titanium oxides not seen by Spitzer \citep{2012A&A...543L...2S} will allow testing of the chemical calculations presented here. The detection of TiO and TiO$_2$ in the cooler region of late-type stars, implies that other small transition metal--bearing molecules such as VO, might be found with sensitive interferometers in the submillimeter wave band \citep{2013A&A...551A.113K}. It is interesting to note that one might prefer the VN/VO$_2$ ratio over the one we chose above, since the VO$_2$ molecule exists in gaseous form at relatively lower temperatures than VO (VO$_2$ peaks at 1600 K, VO at 1850 K; see panel (c) of Fig. \ref{Ti1}), and thus farther in the disk from the host star, a feature that might allow the region to be better resolved, decreasing the chance of stellar contamination. 

\acknowledgements

We thank an anonymous Referee for his useful comments that helped us in improving our manuscript. O. Mousis acknowledges support from CNES. JIL acknowledges support from the JWST program through a grant from NASA Goddard. We thank T. Guillot and R. Hueso for having provided us with their accretion disk model. We thank A. Rajpurohit for his useful comments on stellar temperatures.

\end{document}